\documentclass[12pt]{article}
\usepackage{graphicx}
\usepackage{epstopdf}
\usepackage{dcolumn}
\usepackage{bm}
\usepackage{hyperref}
\usepackage{color}

\begin{document}


\def\a{\alpha}
\def\b{\beta}
\def\c{\varepsilon}
\def\d{\delta}
\def\e{\epsilon}
\def\f{\phi}
\def\g{\gamma}
\def\h{\theta}
\def\k{\kappa}
\def\l{\lambda}
\def\m{\mu}
\def\n{\nu}
\def\p{\psi}
\def\q{\partial}
\def\r{\rho}
\def\s{\sigma}
\def\t{\tau}
\def\u{\upsilon}
\def\v{\varphi}
\def\w{\omega}
\def\x{\xi}
\def\y{\eta}
\def\z{\zeta}
\def\D{\Delta}
\def\G{\Gamma}
\def\H{\Theta}
\def\L{\Lambda}
\def\F{\Phi}
\def\P{\Psi}
\def\S{\Sigma}

\def\o{\over}
\def\beq{\begin{eqnarray}}
\def\eeq{\end{eqnarray}}
\newcommand{\gsim}{ \mathop{}_{\textstyle \sim}^{\textstyle >} }
\newcommand{\lsim}{ \mathop{}_{\textstyle \sim}^{\textstyle <} }
\newcommand{\vev}[1]{ \left\langle {#1} \right\rangle }
\newcommand{\bra}[1]{ \langle {#1} | }
\newcommand{\ket}[1]{ | {#1} \rangle }
\newcommand{\EV}{ {\rm eV} }
\newcommand{\KEV}{ {\rm keV} }
\newcommand{\MEV}{ {\rm MeV} }
\newcommand{\GEV}{ {\rm GeV} }
\newcommand{\TEV}{ {\rm TeV} }
\newcommand{\1}{\mbox{1}\hspace{-0.25em}\mbox{l}}
\newcommand{\headline}[1]{\noindent{\bf #1}}
\def\diag{\mathop{\rm diag}\nolimits}
\def\Spin{\mathop{\rm Spin}}
\def\SO{\mathop{\rm SO}}
\def\O{\mathop{\rm O}}
\def\SU{\mathop{\rm SU}}
\def\U{\mathop{\rm U}}
\def\Sp{\mathop{\rm Sp}}
\def\SL{\mathop{\rm SL}}
\def\tr{\mathop{\rm tr}}
\def\mpl{M_{\rm Pl}}

\def\IJMP{Int.~J.~Mod.~Phys. }
\def\MPL{Mod.~Phys.~Lett. }
\def\NP{Nucl.~Phys. }
\def\PL{Phys.~Lett. }
\def\PR{Phys.~Rev. }
\def\PRL{Phys.~Rev.~Lett. }
\def\PTP{Prog.~Theor.~Phys. }
\def\ZP{Z.~Phys. }

\def\dd{\mathrm{d}}
\def\ff{\mathrm{f}}
\def\BH{{\rm BH}}
\def\inf{{\rm inf}}
\def\ev{{\rm evap}}
\def\eq{{\rm eq}}
\def\SM{{\rm sm}}
\def\Mpl{M_{\rm Pl}}
\def\GeV{{\rm GeV}}
\newcommand{\Red}[1]{\textcolor{red}{#1}}

\begin{titlepage}
\begin{center}

\hfill IPMU-14-0298 \\
\hfill ICRR-report-691-2014-17 \\
\hfill \today

\vspace{1.5cm}
{{\large\bf 
$R$-symmetric Axion/Natural Inflation in Supergravity via Deformed Moduli Dynamics
}
}

\vspace{2.0cm}
{\bf Keisuke Harigaya}$^{(a)}$,
{\bf Masahiro Ibe}$^{(a, b)}$,
and 
{\bf Tsutomu T. Yanagida}$^{(a)}$

\vspace{1.0cm}
{\it
$^{(a)}${Kavli IPMU (WPI), University of Tokyo, Kashiwa, Chiba 277-8583, Japan} \\

$^{(b)}${ICRR, University of Tokyo, Kashiwa, Chiba 277-8582, Japan}
}

\vspace{2.0cm}
\abstract{
We construct a natural inflation model in supergravity where 
the inflaton is identified with a modulus field possessing 
a shift symmetry.
The superpotential for the inflaton
is generated by meson condensation due to strong dynamics
with deformed moduli constraints.
In contrast to models based on gaugino condensation, 
the inflaton potential is generated without $R$-symmetry breaking and hence does not depend on the gravitino mass.
Thus, our model is compatible with low scale supersymmetry.
}
\end{center}
\end{titlepage}
\setcounter{footnote}{0}

\section{Introduction}

Slow-roll inflation~\cite{Linde:1981mu,Albrecht:1982wi} is now a standard paradigm in the modern cosmology.
It not only solves the flatness problem and the horizon problem~\cite{Guth:1980zm,Kazanas:1980tx}, but it also explains 
the origin of the large scale structure of the universe~\cite{Mukhanov:1981xt,Hawking:1982cz,Starobinsky:1982ee,Guth:1982ec,Bardeen:1983qw}.
This paradigm has been supported by precise measurements of the cosmic microwave background (CMB)~\cite{Smoot:1992td,Hinshaw:2012aka,Ade:2013uln}.

After the announcement by the BICEP2 experiment on the B-mode polalization~\cite{Ade:2014xna}, 
models with larger inflaton field values than the Planck scale are drawing much attention due to 
the so-called Lyth bound\,\cite{Lyth:1996im}.%
\footnote{Models with large inflaton field value are free from the initial condition problem~\cite{Linde:2005ht}.
}
Such a large field value seems inconsistent with the conventional view of the
field theoretic description as an effective theory which is believed to be at the best given by 
a series expansion of fields with higher dimensional operators suppressed by the Planck scale.
In other words, in large field inflation models, any higher dimensional terms of the inflaton potential 
should be somehow under control.

The best way to understand such strict control on the inflaton potential would be a shift symmetry of the inflaton~\cite{Kawasaki:2000yn}.
Interestingly, such a candidate of the inflaton with a  shift symmetry is often provided in string theories 
as a modulus~\cite{Svrcek:2006yi}. 
We refer to the modulus as an axion, although it is not the QCD axion which solves the strong CP problem~\cite{Peccei:1977hh,Weinberg:1977ma,Wilczek:1977pj}.
Once we identify the axion as the inflaton, the next task is to 
generate a potential of the axion.
As a caveat, in the situation where the shift symmetry holds at the tree level and is broken by quantum effects, 
as is often the case with axions in superstring theories,
the superpotential of the axion, and hence, the axion potential, is generated only by non-perturbative effects~\cite{Grisaru:1979wc}.
Thus, model construction often requires strong gauge dynamics to generate the axion potential.

Along this line, natural inflation models in supergravity have been constructed~\cite{Freese:1990rb,Adams:1992bn,Kim:2004rp,Dimopoulos:2005ac,Czerny:2014xja,Dine:2014hwa,Yonekura:2014oja}.%
\footnote{
For inflation models other than natural inflation where inflaton potentials are generate dynamically, see Refs.~\cite{Dimopoulos:1997fv,Izawa:1997df,Izawa:1997jc,Harigaya:2012pg,Harigaya:2014wta}.
}
As a common feature of these models, the axion potential originates from gaugino condensation in
strongly coupled gauge theories.
As a result, the energy scale of the axion potential is proportional to
the scale of $R$ symmetry breaking,
i.e, the gravitino mass.
Thus, to explain the magnitude of cosmic perturbations, the gravitino mass is required to be 
as large as $10^{13}$ GeV, which is incompatible with low scale supersymmetry breaking.%
\footnote{
In Refs~\cite{Kallosh:2014vja,Harigaya:2014eta}, natural inflation models consistent with low scale supersymmetry breaking are proposed, although the shift symmetry breaking is simply given by tree-level superpotentials.
}

In this letter, we propose to make use of meson condensation by strong dynamics with deformed moduli constraints
to 
generate the superpotential of the axion/inflaton field.%
\footnote{The idea of generating axion potential by the meson condensation is suggested in Ref.~\cite{Henning:2014dha}.
}
As we will show, the model possesses an $R$-symmetry and the inflaton potential does not
depend on the gravitino mass.
Thus, our model is compatible with low energy supersymmetry breaking.

\section{Inflaton potential from meson condensation}

\subsubsection*{Dynamical Sector}

Let us begin with a brief review on a supersymmetric $SP(N_c)$ gauge theory 
with $2(N_c +1)$ chiral superfields in the fundamental 
representation, $Q^i(i =1 \cdots 2(N_c + 1))$.%
\footnote{In our convention, $SP(1)$ is equivalent to $SU(2)$.}
The vacuum structure of classical flat directions, i.e. $(N_c + 1)(2N_c + 1)$ meson fields,
\begin{eqnarray}
M^{ij} \propto Q^i Q^j \ ,
\end{eqnarray}
is deformed non-perturbatively.
The vacuum expectation values (VEVs) of the meson fields obey 
the so-called deformed quantum moduli constraint~\cite{Seiberg:1994bz};
\begin{eqnarray}
Pf^{(N_c+1)}(M^{ij}) = \Lambda^{N_c + 1}\ .
\label{eq:constraint}
\end{eqnarray}
Here, $\Lambda$ denotes the dynamical scale of the $SP(N_c)$ gauge interaction
and $Pf (\cdots)$ denotes the Pfaffian.%
\footnote{%
We define the Pfaffian of a $2n\times 2n$ antisymmetric matrix, 
$Pf^{(n)}$, so that the symplectic form $J$,
where $J = \1_n\otimes i\s_2$ with $\1_n$ being the $n\times n$ 
unit matrix and $\s_2$ being the second Pauli matrix, satisfies $Pf^{(n)}(J)=1$.}
We have normalized the meson fields $M^{ij}$ so that they have a mass dimension one.
As is clear from Eq.\,(\ref{eq:constraint}),  some of the mesons condensate at the vacuum.

\subsubsection*{Axion}
Next, let us introduce an axion chiral multiplet $T$ 
which couples to the above gauge dynamics via the gauge kinetic function.
Later on, we will identify the imaginary part of the axion multiplet $T$ with the inflaton.
To be concrete, we assume that the K\"ahler potential of the axion multiplet is given by%
\footnote{
Here, we have chosen the origin of $T$
so that  the K\"ahler potential does not have a linear term $T+T^\dagger$.
}
\begin{eqnarray}
K = K(T+T^\dagger) = \frac{1}{2}(T+T^\dagger)^2 + \cdots\ , 
\end{eqnarray}
where the ellipses denotes higher dimensional terms.
Here, we have assumed that the K\"ahler potential has a 
shift symmetry,
$T \to T + i \alpha$, with $\alpha$ being a real number.
We also assume that the axion multiplet appears in the gauge kinetic function of the $SP(N_c)$ gauge multiplet,
\begin{eqnarray}
{\cal L}_{\rm gauge}= \frac{1}{4} \int {\rm d}^2 \theta \left(\frac{1}{g^2} + \frac{T}{ 8\pi^2 f_a}\right) W^\alpha W_\alpha + {\rm h.c.} .
\end{eqnarray}
where a dimensionful constant
$f_a$
denotes the ``decay constant" 
which depends on the origin of the axion multiplet.
We assume that this coupling is the dominant contribution to the shift symmetry breaking of the axion.

In our argument, instead of specifying the origin of the axion multiplet, 
we simply assume that the value of $f_a$ is at around the so-called string scale, i.e. $M_{\rm str} \simeq 10^{17}$\,GeV,
which is expected in the case of string axions\,\cite{Svrcek:2006yi}.
Through the coupling to the gauge kinetic term, the shift symmetry is broken by the non-perturbative effects of the $SP(N_c)$ dynamics.

\subsubsection*{STEP1}
In the presence of the axion multiplet in the kinetic function, the effective dynamical scale depends 
on the axion field, i.e.,
\begin{eqnarray}
 \Lambda_{\rm eff} (T) = \Lambda \exp\left[-{\frac{1}{2(N_c+1)}\frac{T}{f_a}}\right]\ .
\end{eqnarray}
Accordingly, the above meson condensation in Eq.\,(\ref{eq:constraint}) also depends on the axion multiplet, i.e.
\begin{eqnarray}
Pf^{(N_c+1)}(M^{ij}) = \Lambda_{\rm eff}^{N_c + 1}(T)\ .
\label{eq:constraint2}
\end{eqnarray}
It should be emphasized here that mere condensation of the mesons does not lead to a non-trivial
potential of the axion multiplet, although the meson condensation scale depends on the axion multiplet.
This feature should be contrasted with the axion potential generation via the gaugino 
condensation, where the condensation leads to a non-trivial potential of the axion multiplet.

\subsubsection*{STEP2}
To generate a non-trivial axion potential, let us introduce $(N_c + 1)(2N_c + 1)$
singlet fields, $X^{ij}= - X^{ji}$, which couple to the fundamental fields $Q^i$
in the same way with the model of dynamical supersymmetry breaking developed in Refs.~\cite{Izawa:1996pk,Intriligator:1996pu};
\begin{eqnarray}
W = \sum_{i>j,k>l}\lambda_{ij,kl} X^{ij} Q^k Q^l\ .
\label{eq:tree1}
\end{eqnarray}
To make our analysis simple, 
we hereafter assume that the above superpotential possesses 
a global $SP(2(N_c + 1))$ symmetry out of the maximal flavor $SU(2(N_c+1))$ symmetry,
and that the $SP(2(N_c+1))$ singlet direction, $X^{ij} \propto J^{ij}$,
has the smallest coupling to the quarks, i.e.,%
\footnote{The following arguments can be extended to
generic cases as done in Ref.\,\cite{Harigaya:2014wta}.}
\begin{eqnarray}
\lambda_{ij,kl}  = \lambda''J_{ik}J_{jl} + \left(\lambda'- \frac{\lambda''}{2(N_c +1)} \right)J_{ij}J_{kl}\ , \quad (|\lambda'| < |\lambda''|)\ .
\end{eqnarray}

Below the dynamical scale, the tree-level interactions lead
to effective couplings between the mesons and the singlets,
\begin{eqnarray}
W_{\rm eff} \simeq  \sum_{i>j,k>l} \lambda_{ij,kl} \Lambda_{\rm eff}(T) X^{ij} M^{kl}\ ,
\end{eqnarray}
where the mesons are subject to the deformed constraint in Eq.\,(\ref{eq:constraint2}).%
\footnote{We may consider the deformed moduli constraint as a consequence
of equations of motions of heavy states such as glueball supermultiplet of $SP(N_c)$.
Following arguments are not significantly altered even when we treat the 
deformed moduli constraints as the equation of motion of heavy states.
}
In this effective theory, we see that all the meson fields and the singlets get massive 
at around the VEVs of the mesons,
\begin{eqnarray}
 M^{ij} = \Lambda_{\rm eff}(T) \times J^{ij}\ ,
 \label{eq:mesonC}
\end{eqnarray}
except for the singlet which corresponds to the 
global $SP(2(N_c+1))$ singlet.%
\footnote{One of the meson obtains a mass of $O(\Lambda_{\rm eff})$ due to the 
deformed constraint.}
By inserting this solution to the effective potential, we obtain the effective superpotential
of the remaining singlet field,
\begin{eqnarray}
W_{\rm eff} \simeq \lambda \,\Lambda_{\rm eff}(T)^2  X
\simeq \lambda \,\Lambda^2 \, e^{-\frac{1}{(N_c+1)}\frac{T}{f_a}}X \ ,
\label{eq:Weff}
\end{eqnarray}
after integrating out other heavier mesons and singlets.%
\footnote{Mixing between the axion and the mesons is suppressed by $\Lambda_{\rm eff}/ \left(\left(N_c +1\right) f_a\right)$ and hence negligible~\cite{Henning:2014dha}.}
Here, we have defined 
\begin{eqnarray}
X &=& \frac{1} {\sqrt{N_c + 1}}\sum_{i>j} J^{ij} X_{ij} \ ,\\
\lambda &=& 2 \lambda' \left(N_c +1\right)^{3/2}.
\end{eqnarray}
As a result, we find that the supersymmetry is broken for a given value of $T$, which leads 
to a nontrivial potential of the axion field,
\begin{eqnarray}
V_{\rm eff} \simeq | \lambda|^2 \Lambda^4 e^{-\frac{1}{(N_c+1)}\frac{T+ T^\dagger}{f_a}} \ ,
\label{eq:runaway}
\end{eqnarray}
where we have set $X = 0$.%
\footnote{Here, it should be noted that the scalar component of $X$ is stabilized to $X = 0$ due to
a large positive mass of the scalar component, $\Delta m_X^2 \sim \lambda^4 H_{\rm inf}^2 \Mpl^2/\Lambda_{\rm eff}^2$, generated
by perturbative corrections~\cite{Chacko:1998si}.
In the limit of a small dynamical scale, the mass of the scalar component is far larger than the Hubble scale and the scalar component decouples during inflation.
Thus, $X$ can be identified with a nilpotent chiral superfield discussed in Ref.~\cite{Ferrara:2014kva}.
In our model, however, $\Delta m_X^2$ vanishes and the scalar component becomes light after inflation.
The chiral multiplet $X$ becomes a mass partner of the inflaton and is relevant for the decay of the inflaton.
}
Unfortunately, however, the imaginary part of the scalar component of $T$, the axion field, remains flat, and hence,
this dynamics does not lead to the model of natural inflation.

\subsubsection*{STEP3}
The above failure can be traced back to the remaining shift-symmetry in the effective 
potential in Eq.\,(\ref{eq:Weff}) under which $X$ rotates to 
absorb the shift of $T$.%
\footnote{Accordingly, the fundamental fields also rotate under the remaining symmetry
which makes the original shift symmetry free of the anomaly, and hence,
one linear combination of the phases remains as a massless axion.}
Therefore, to generates non-trivial potential for the imaginary part of the axion,
we are lead to add a linear term of $X$ which breaks the remaining shift-symmetry explicitly,
\begin{eqnarray}
{\mit\Delta} W = - \mu^2 X ,
\label{eq:explicit}
\end{eqnarray}
where $\mu$ is a dimensionful parameter.%
\footnote{If we extend the definition of the shift symmetry so that
$X$ rotates non-trivially, we may add a term ${\mit \Delta }W' = e^{c T} X$
with an appropriate coefficient $c$, which is in general broken 
by the anomaly of the $SP(N_c)$ gauge interaction.
With such a term, we obtain a different inflaton potential, although we do not
pursue this possibility in this paper.
}
We note that this term is consistent with the $R$-symmetry which we discuss later.
In the presence of the breaking term, the above dynamics leads to the effective potential,
\begin{eqnarray}
W &=&  \lambda \,\Lambda^2 \left(e^{\frac{1}{(N_c+1)}\frac{T}{f_a}}- \tilde{\mu}^2 \right) X\ , \\
\tilde{\mu}^2  &=& \frac{\mu^2}{ \lambda \,\Lambda^2}\ .
\label{eq:Weff2}
\end{eqnarray}
In the followings, we take a phase convention of $X$ so that $\tilde\mu$ is real and positive valued.
As a result, we obtain an axion potential,
\begin{eqnarray}
V_{\rm eff} \simeq  | \lambda|^2 \Lambda^4 \left(
(e^{\sqrt{2}\frac{\tau}{(N_c + 1)f_a}} + \tilde{\mu}^4)
-2\tilde{\mu}^2e^{
\frac{\tau}{\sqrt{2}(N_c +1)f_a}} 
\cos\left[ 
 \frac{\phi}{\sqrt{2}(N_c+1)f_a} 
\right]
\right)\ ,
\end{eqnarray}
which lifts up the imaginary part of the axion field.
In the above expression, we have
decomposed the axion field into 
\begin{eqnarray}
T = \frac{1}{\sqrt{2}} (\tau + i \phi) \ .
\end{eqnarray}

It should be noted that unlike the model of dynamical supersymmetry breaking model in 
~\cite{Izawa:1996pk,Intriligator:1996pu},
the model does not break supersymmetry spontaneously due to the presence of $T$, 
where the supersymmetry vacuum is at
\begin{eqnarray}
e^{\frac{\tau}{\sqrt{2}(N_c + 1)f_a}} \simeq \tilde{\mu}^2\ , \quad \phi = 0 \ .
\label{eq:SUSYvac}
\end{eqnarray}
At around this vacuum, both the axion and its real field counterpart obtain the same mass,
\begin{eqnarray}
m^2 = \frac{\mu^4}{(N_c+1)^2f_{a}^2}\ .
\end{eqnarray}
It should be also noted that the resultant scalar potential does not show the runaway
behavior as seen in Eq.\,(\ref{eq:runaway}).

Now, let us assume that the real part of the axion field is fixed to its supersymmetric vacuum value,
while allowing the axion field being away from its vacuum value, i.e. $\phi\neq 0$.
In this case, the superpotential Eq.\,(\ref{eq:Weff2}) is reduced to
\begin{eqnarray}
W =  \sqrt{2}\mu^2 \left(e^{\frac{1}{(N_c+1)}\frac{i\phi}{\sqrt{2}f_a}} - 1\right) X\ ,
\end{eqnarray}
where $\phi$ should be understood as not a chiral field but a constant.
It should be noted that our model has the same structure as the ``Model 1" in Ref.\,\cite{Kallosh:2014vja},
and hence, our model provides an ultraviolet completion to their model. 

As a result, along the lines of the chaotic inflation model with shift symmetry in \cite{Kawasaki:2000yn,Kallosh:2014vja},
the axion field obtains a nontrivial potential through the $F$-term contribution of $X$
which leads to
\begin{eqnarray}
V_{\rm eff} \simeq  2\mu^4\left(1 -
\cos\left[ 
\frac{\phi}{f_{\rm eff}} 
\right]
\right)\ .
\end{eqnarray}
Here, we have defined an effective decay constant,
\begin{eqnarray}
 f_{\rm eff} =  \sqrt{2}\, (N_c + 1)f_a \ .
\end{eqnarray}
It should be noted that the effective decay constant is required to be larger than 
the Planck scale to satisfy the slow-roll conditions in natural inflation.
For $f_a = O(M_{\rm str}) = O(10^{17})$ GeV,
the effective decay constant is larger 
than the Planck scale if $N_c = O(10)$.%
\footnote{
Enhancement of the effective decay constant in the inflaton potential by a large $N_c$ is pointed out in Refs.~\cite{Dine:2014hwa,Yonekura:2014oja,Kaloper:2011jz}.
In the view point of the ``Phase Locking Mechanism" proposed in Refs.~\cite{Harigaya:2014eta},
the enhanced decay constant is understood by hierarchical charges between the phase of $X$ and $\phi$ under the remaining shift symmetry discussed at the beginning of this subsection, and the breaking of the remaining shift symmetry by the superpotential term in Eq.~(\ref{eq:explicit}).
}

In this way, we find that the model with meson condensation leads to the inflaton potential
which is appropriate for natural inflation.
For recent discussion on the consistency of natural inflation with CMB data, we refer e.g. Ref.\,\cite{Freese:2014nla}.

\subsubsection*{Required Tuning}
We clarify how feasible it is to assume 
that the real part of  $\tau$ is fixed to a desirable position in Eq.\,(\ref{eq:SUSYvac}). 
For that purpose, let us first estimate the mass of the real part of the axion
around the field value in Eq.\,(\ref{eq:SUSYvac}), which is given by,
\begin{eqnarray}
\label{eq:Tmass}
m_\tau^2 = \frac{4 \mu^4}{f_{\rm eff}^2} \ll 
\frac{4 \mu^4}{f_{\rm eff}^2} \frac{\phi^2}{M_{\rm PL}^2} \simeq H_{\rm inf}^2\ ,
\end{eqnarray}
where $H_{\rm inf}$ denotes the Hubble parameter during inflation.
Thus, the real part of the axion is not fixed by the superpotential coupling to $X$,
and hence, we need to have the axion coupling to 
$X$ in the K\"ahler potential which is in general given by,
\begin{eqnarray}
{\mit \Delta}K = \frac{X^\dagger X}{M_{\rm PL}^2}
\left( \sqrt{2}c_1 M_{PL}  (T+T^\dagger)  +  c_2 (T+T^\dagger)^2/2 + \cdots\right)\ ,
\end{eqnarray}
where $c_{1,2}$ are $O(1)$ coefficients.
With these terms, the real part of the axion field is fixed to
\begin{eqnarray}
\tau_* \simeq \frac{c_1}{1-c_2} M_{\rm PL}\ ,
\end{eqnarray}
where we have assumed $|c_1|\ll 1$, for simplicity.
In general, this field value is expected to be far away from the vacuum position in Eq.\,(\ref{eq:SUSYvac}).

If the real  part of the axion field is fixed at far away from the 
vacuum position, the axion stays at $\tau_*$, and never
goes back to the vacuum position after inflation since the effective mass 
of the real part of the axion around $\tau_*$ is much larger than the one in Eq.\,(\ref{eq:Tmass}).
Hence, inflation never ends due to the non-vanishing potential energy
at $\tau_*$ even for $\phi = 0$, i.e.
\begin{eqnarray}
V_{\rm eff} \simeq  | \lambda|^2 \Lambda^4 \left(
e^{\frac{\tau}{\sqrt{2}(N_c + 1)f_a}} - \tilde{\mu}^2
\right)^2 \neq 0\ .
\end{eqnarray}
Thus, in order to avoid this problem, we need to tune the value of $\mu$,
so that 
\begin{eqnarray}
e^{\frac{\tau_*}{\sqrt{2}(N_c + 1)f_a}} = \tilde{\mu}^2(1 + \delta)\ , \quad (\delta \ll 1)\ .
\label{eq:SUSYvac2}
\end{eqnarray}
With this tuning, the inflaton potential along $\tau_*$ is given by
\begin{eqnarray}
V_{\rm eff} \simeq  2\mu^4(1+\delta)\left(1 + 
\frac{\delta^2}{2} -
\cos\left[ 
 \frac{\phi}{\sqrt{2}(N_c + 1)f_a} 
\right]
\right)\ .
\end{eqnarray}
By remembering that the mass of $\tau$ around $\tau_*$ is given by,
\begin{eqnarray}
 m_{\tau_*}^2(\phi) \simeq \frac{(1-c_2) V_{\rm eff}(\phi)}{M_{\rm PL}^2}\ ,
\end{eqnarray}
we find that the axion field goes back to the vacuum position well after inflation, i.e. $\phi\simeq 0$ 
as long as
\begin{eqnarray}
m_\tau^2  > m_{\tau_*}^2(\phi \simeq 0) \simeq \frac{(1-c_2) \mu^4 \delta^2}{M_{\rm PL}^2} \ .
\end{eqnarray}
To satisfy the above condition, we find that we need tuning between parameters,
\begin{eqnarray}
\delta < \frac{4 M_{\rm PL}^2}{(1-c_2)f^2_{\rm eff}}\ .
\label{eq:finetuning}
\end{eqnarray}

\subsubsection*{$R$-symmetry}
Finally, we note that the $R$-symmetry is preserved in our model.
The $R$-charge assignment is $X(2)$, $Q^i(0)$ and $T(0)$.
The $R$-symmetry is free from the gauge anomaly of the $SP(N_c)$, and hence not explicitly broken by the strong dynamics of the $SP(N_c)$ gauge theory.
Also, since the scalar component of $X$ is fixed to its origin, the $R$-symmetry is also not spontaneously broken.
Thus, the inflaton sector does not break the $R$-symmetry, and hence the inflation scale is not related with the gravitino mass.
Our model is compatible with low scale supersymmetry.

We stress that the $R$-symmetry is important for stable inflaton dynamics.
If the $R$-symmetry is broken during inflation, the negative contribution to the inflaton potential is significant 
and the inflaton may be destablized toward far from the origin. 
In our model, since the $R$-symmetry is preserved during inflation,
the negative contribution is absent.

We have made use of meson condensation to generate the inflaton potential.
As is pointed out in Ref.~\cite{Izawa:1998dv}, the mechanism can be applied to moduli fixing.
Since moduli are fixed in an $R$ invariant way, masses of moduli can be far larger than the gravitino mass.
Thus, moduli fixing by meson condensation is free from destabilization of moduli during inflation by Hubble induced potentials.

\section{Summary and discussion}
In this letter, we have proposed a natural inflation model in supergravity where 
the axion potential is generated by meson condensation due to strong dynamics with deformed moduli constraints..
In contrast to models based on gaugino condensation, 
our model possesses an unbroken $R$-symmetry and hence the inflaton potential does not
depend on the gravitino mass.
Thus, our model is compatible with low scale supersymmetry. 

In the above analysis, we have assumed one axion field.
It is easy to extend our model to multi-axion cases.
For example, let us consider two axions $T$ and $S$.
We couple them to two gauge theories via gauge kinetic functions and assume that gauge theories are
in meson condensation phases.
By fixing mesons in the same way as the above analysis, we obtain the effective super potential,
\begin{eqnarray}
W_{\rm eff} =
X \Lambda^2 \left( {\rm exp}\left[ \frac{T}{f_T} + \frac{S}{f_S} \right]  - \tilde{\mu}^2 \right)+
X' \Lambda^{'2} \left( {\rm exp}\left[ \frac{T}{f'_T} + \frac{S}{f'_S} \right]  - \tilde{\mu}^{'2} \right),
\end{eqnarray}
where $X$ and $X'$ are singlel fields corresponding to that in Eq.~(\ref{eq:Weff}) for two gauge theories, and
$\Lambda^{(')}$, $f_T^{(')}$, $f_S^{(')}$ and $\tilde{\mu}^{(')}$ are constants.
If $f_T/ f_S \simeq f'_T/f'_S$, a linear combination of $T$ and $S$ works as an inflaton with a effective decay constant much larger than $f_T^{(')}$ and $f_S^{(')}$~\cite{Kim:2004rp}.

We have assumed the global $SP\left(2\left(N_c+1\right)\right)$ symmetry to simplify our analysis.
Without the symmetry, the VEVs of mesons are not given by Eq.~(\ref{eq:mesonC}), but generic ones which depend on constants $\lambda$s and $\mu^2$s. 
After integrating out heavy mesons and singlets, the effective superpotential is given by Eq.~(\ref{eq:Weff}), but $\lambda$, $\tilde{\mu}^2$ and $\Lambda$ in general depends on $T$.
As a result, the inflaton potential is not given by a simple cosine form.
It is interesting if deviation from the cosine form is observed.

Let us comment on decay of the inflaton.
The inflaton does not possess any charges under some linearly realized symmetry.
Thus, the inflaton in general decays into standard model particles through its linear terms in the K\"ahler potential or gauge kinetic functions~\cite{Endo:2006qk,Endo:2007ih}.

The inflaton also decays into supersymmetry breaking sector fields, which may lead to the overproduction of gravitinos~\cite{Endo:2006qk,Endo:2007ih,Kawasaki:2006gs,Asaka:2006bv,Dine:2006ii,Endo:2006tf}.
The overproduction can be avoided if masses of supersymmetry breaking sector fields are large enough, 
so that the decay mode is kinematically forbidden~\cite{Nakayama:2012hy}.%
\footnote{
The inflaton also decays into a pair of gravitinos through the mixing between the inflaton and a scalar component of the supersymmetry breaking field.
This decay mode is suppressed if the supersymmetry breaking field is weakly coupled. For detailed discussion, see Ref.~\cite{Nakayama:2012hy}.
}
As is discussed in Ref.~\cite{Harigaya:2014roa},
the supersymmetry breaking scale may have a lower bound.
It may be interesting that the large supersymmetry breaking scale assumed in the pure gravity mediation~\cite{Ibe:2006de,Ibe:2011aa,Ibe:2012hu} is naturally explained in this way.

\section*{Acknowledgements}
T.T.Y thanks for discussion with Brian~Henning, John~Kehayias, Hitoshi~Murayama and David~Pinner.
The authors thank Taizan~Watari for useful comments. 
This work is supported by Grant-in-Aid for Scientific Research from the
Ministry of Education, Science, Sports, and Culture (MEXT), Japan,
No.\ 26104009 and 26287039 (T.\,T.\,Y.) and No.\ 24740151 and 25105011 (M.\,I.),
from the Japan Society for the Promotion of Science (JSPS), No. 26287039 (M.I.),
as well as by the World Premier
International Research Center Initiative (WPI), MEXT, Japan.
The work of K.\,H.\ is supported in part by a Research Fellowship
for Young Scientists from the Japan Society for the Promotion of Science (JSPS).

\end{document}